\documentclass[a4paper,aps,prx,twocolumn,floatfix,superscriptaddress,notitlepage,usenames,dvipsnames,svgnames,table,footinbib,longbibliography]{revtex4-1}
\pdfoutput=1
\usepackage[utf8]{inputenc}
\usepackage[T1]{fontenc}
\usepackage{graphicx}
\usepackage{grffile}
\usepackage{wrapfig}
\usepackage{rotating}
\usepackage[normalem]{ulem}
\usepackage{amsmath}
\usepackage{textcomp}
\usepackage{amssymb}
\usepackage{capt-of}

\usepackage{parskip}
\usepackage{mathrsfs}
\usepackage[margin= 0.75in]{geometry}
\usepackage[braket, qm]{qcircuit}

\usepackage[english]{babel}
\usepackage{letltxmacro}
\LetLtxMacro{\ORIGselectlanguage}{\selectlanguage}
\makeatletter
\DeclareRobustCommand{\selectlanguage}[1]{%
  \@ifundefined{alias@\string#1}
    {\ORIGselectlanguage{#1}}
    {\begingroup\edef\x{\endgroup
      \noexpand\ORIGselectlanguage{\@nameuse{alias@#1}}}\x}%
}
\newcommand{\definelanguagealias}[2]{%
  \@namedef{alias@#1}{#2}%
}
\makeatother
\definelanguagealias{en}{english}
\definelanguagealias{eng}{english}

\usepackage{bbm}
\usepackage{etoolbox}
\usepackage{tcolorbox} 
\usepackage{tikz}
\usepackage{amsthm}
\usepackage{amssymb}
\usetikzlibrary{arrows.meta}
\usepackage{mathtools}

\usepackage{bm}
\usepackage{dsfont}
\usepackage[font=small,labelfont=bf,justification=RaggedRight,format=plain]{caption}
\usepackage{subcaption}

\usepackage{thmtools}
\usepackage{thm-restate}

\newtheorem{proposition}{Proposition}

\usepackage{microtype}
\usepackage{dsfont}
\usepackage{tensor}

\usepackage{physics}
\usepackage{mathrsfs}
\usepackage{mathtools}
\usepackage{fixltx2e}
\usepackage{relsize}

\usepackage{float}
\usepackage[Export]{adjustbox} 
\adjustboxset{max size={0.9\linewidth}{0.9\paperheight}}

\usepackage[font=small,labelfont=bf,justification=RaggedRight,format=plain]{caption}
\usepackage{subcaption}

\DeclarePairedDelimiterX\phys[2]{\langle}{\rangle}{#1 \delimsize\vert\mathopen{} #2}

\theoremstyle{remark}

\definecolor{blue-violet}{rgb}{0.54, 0.17, 0.89}
\usepackage{hyperref}
\hypersetup{
 pdfauthor={Paolo Zanardi et al},
 pdftitle={Mutual averaged non-commutativity of quantum observable algebras}
 pdflang={English},colorlinks=true,linkcolor=RubineRed,citecolor=blue-violet,urlcolor=Cerulean} 
\usepackage[capitalise]{cleveref}

\begin{document}

\title{Mutual averaged non-commutativity of quantum operator algebras}

\author{Paolo Zanardi}
\affiliation{Department of Physics and Astronomy, and Center for Quantum Information Science and Technology, University of Southern California, Los Angeles, California 90089-0484, USA}
\affiliation{Department of Mathematics, University of Southern California, Los Angeles, California 90089-2532, USA}

\date{\today}
\begin{abstract}
We introduce an elementary  measure of non-commutativity between two algebras of quantum operators acting on the same Hilbert space.  This quantity,  which we call Mutual Averaged Non-commutativity (MAN), is a simple generalization of a type of averaged Out-of-Time-Order-Correlators used in the study of quantum scrambling and chaos.
MAN is defined by a Haar averaged squared norm of a  commutator and for some types of algebras is  manifestly of entropic nature.
 In particular,  when the two algebras coincide  the corresponding self-MAN can be fully computed in terms of the structural data of the associated  Hilbert space decomposition. 
 Properties and bounds of MAN are established in general and several concrete examples are discussed.  Remarkably,  for an important class of algebras, --which includes factors and maximal abelian ones --  MAN can be expressed in the terms of the algebras projections CP-maps.  Assuming that the latter can be enacted as physical processes,  one can devise operational  protocols  to directly estimate the MAN of a pair of algebras.   
 

\end{abstract}

\maketitle
\section{Introduction}\label{sec:intro}
Non-commutativity of physical observables is one of the key features of quantum theory.  In one way or another it provides the mathematical underpinnings  and conceptual foundation of distinctive  quantum phenomena as uncertainty relations,  quantum coherence and quantum entanglement.  Because of this non-commutativity,  can also be viewed as a resource  for quantum information processing and computing \cite{nielsen_quantum_2010}. 
As a matter of fact,  a system whose observable algebra is abelian is {\em{fundamentally}} classical in nature  \footnote{This can be regarded  as the physical content the Gelfand  representation theorem, see e.g.,  Arveson, W. (1981). An Invitation to $C^*$-Algebras. Springer-Verlag.}.
 
 More recently,  non-commutativity,  as measured by various Out-of-Time-Order-Correlators (OTOCs),  has been shown to provide a powerful diagnostic tool to study  quantum information scrambling \cite{larkin1969quasiclassical,kitaev_simple_2015,MaldacenaChaos2016,PhysRevLett.115.131603,PolchinskiSYK2016,PhysRevLett.117.091602,MezeiChaos2017,Roberts2017Chaos}.
Quantum dynamics makes localized observables to spread  spatially along with correlations and information; this process is witnessed,  and limited by the temporal growth of a commutator \cite{bravyi_lieb-robinson_2006}.

In Refs.  \cite{Zanardi-GAAC-2021, A-OTOC-Faidon-2022} the notion of scrambling has been upgraded to  the level of full algebras of observables.
The corresponding OTOC (dubbed $\cal A$-OTOC) is a generalization of the averaged bipartite OTOC discussed in
\cite{styliaris_information_2020} and it is based on norms of commutators averaged over suitable  groups of unitaries. 
In this way a unification of operator entanglement \cite{zanardi2001entanglement} and coherence generating power \cite{zanardiCoherencegeneratingPowerQuantum2017} is achieved and extended to finite temperature and to open dynamics as well \cite{Namit-BROTOC-2021,Open-OTOC-new}.  Very recently the  short-time behavior of the  $\cal A$-OTOC has been employed in  \cite{ZLD-mereology} to suggest a minimal scrambling approach to quantum mereology \cite{carroll-mereology-2021}.

In this paper,  by generalizing the $\cal A$-OTOC approach,  we introduce a quantitative,   elementary and {\em{analytically viable}},  way to measures the amount of non-commutativity between two subalgebras of operators in a Hilbert space.   Here,  one has two (possibly) distinct and not-isomorphic algebras $\cal A$ and $\cal B$ and considering the average of the norm of commutators between pairs of unitaries  $U \in{\cal A}$ and $V\in{\cal B}.$   
When the second algebra is the unitary image of the commutant of the first one recovers the $\cal A$-OTOC  \cite{Zanardi-GAAC-2021, A-OTOC-Faidon-2022}.

This paper is structured as follows.
In Sect. II the basic definition and properties of our non-commutativity measure are discussed and significant examples are provided.  In Sect. III we analyze  the case ${\cal A}={\cal B}$ where general elegant results can be established.  In Sect. IV we  focus on a special, but physically important class, of algebras for which
our measure admits a direct operational interpretation.  Sect VI contains the conclusions.  Finally, 
the Appendix contains the basic definitions and notation concerning the algebras used in this paper as well as the proofs of all of the Propositions.

 \section{MAN}\label{sec:MAN}
Let us begin by defining the basic objects of our paper.

 {\bf{Definition}}  Given two  \emph{hermitian-closed  unital  algebras}   ${\cal A},{\cal B}\subset L({\cal H})$  ($d:=\mathrm{dim} {\cal H}$) their
 {Mutual Averaged Non-commutativity (MAN)} is given by
\begin{align}\label{MAN} 
S({\cal A}: {\cal B})= \frac{1}{2d}\, \mathbb{E}_{X\in{\cal A},\,Y\in{\cal B}}\left[\|[X,\,Y]\|_2^2\right].
\end{align}
where the expectation $ \mathbb{E}$ is with respect to the Haar measure of the unitaries $X$ ($Y$) of $\cal A$ ($\cal B$).  
In words: MAN measures how one algebra fails to be included in the commutant of the other.
\vskip 0.2truecm
It is also useful, as it will become clear later,  to define a logarithmic version of the MAN in  Eq.~(\ref{MAN}) 
\begin{align}\label{log-MAN}
S_2({\cal A}: {\cal B}):=-\log\left(1-S({\cal A}: {\cal B})\right).
\end{align}
We will sometimes use the notation $S({\cal A}|\,{\cal B}):= S({\cal A}: {\cal B}^\prime)$ (and the analog for log-MAN $S_2$).

A particularly interesting case is provided by the {\em{self-MAN}}  i.e.,  $NC({\cal A}):=S({\cal A}:{\cal A}),$ (and $NC_2$ defined by using $S_2$).
This is a measure of the degree of non-commutativity of the algebra $\cal A.$

In order to establish the  physical meaning of these mathematical definitions let us consider the following simple scenario:
two agents, $A$ and $B$\footnote{Importantly,  $A$ and $B$ are {\em{not}} supposed to be associated with separate spatial locations.}  have access to operations in $\cal A$ and $\cal B$ respectively  as well as to an initial shared state $\rho_0.$  The goal is to transmit information from $A$ to $B$ by encoding it with unitaries drawn from $\cal A$ and to decode it  by performing measurements of observables drawn from $\cal B$.  

Given two unitaries $U, V\in{\cal A} $ one  may consider the distance between the states obtained by acting, with $U$ and $V$ on $\rho_0$ and {\em{restricted}} to the algebra $\cal B$.
\begin{align}\label{eq:d_B}
d_{{\cal B},\rho_0} (U, V):=\sup_X |\mathrm{Tr}\left[X({\cal U}^\dagger(\rho_0)-{\cal W}^\dagger(\rho_0))\right]|,
\end{align}
here the supremum is taken over ${X\in{\cal B}, \|X\|_\infty\le 1}$ and  ${\cal U}(X):= U XU ^\dagger$ (and similarly for $V$).

Identically  vanishing  $d_{{\cal B},\rho_0}$  implies that no information can be transmitted by the described protocol.  Indeed,  in this case  all the encoded states appear identical from the perspective of $B.$  Clearly,  this is the case when 
$S({\cal A}: {\cal B})=0$ i.e.,  ${\cal A}\subset {\cal B}^\prime$ since  all the 
 the unitaries in $\cal A$ are symmetries for the observables in $\cal B.$ 
 
 Our first proposition makes this observation more quantitative and tries to shed some light on the operational meaning of MAN.  It establishes a probabilistic connection between  (\ref{MAN}) and  $d_{{\cal B},\rho_0}$ for Haar randomly chosen $U,V\in{\cal A}$ and arbitrary initial state $\rho_0$.

\begin{proposition}\label{prop: prop-0}
Let $U,  V\in {\cal A}$ be random unitaries,  $\mathrm{Pr}$ denote the probability with respect to the Haar-distribution of $U$ and $V.$ 
\begin{align}\label{eq:Markov}
\mathrm{Pr}\left[ \sup_{\rho_0} d_{{\cal B},\rho_0}(U, V)\ge \epsilon  \right]\le \frac{ d\, c({\cal B})  }{\epsilon}\,S^{\frac{1}{2}}({\cal A}: {\cal B})
\end{align}
where $c({\cal B})$ depends just on  $\cal B.$ 
\end{proposition}
Because of the $d$ prefactor,  the  bound ({\ref{eq:Markov}) may be  fairly weak,  or even trivial,  for large $d$.
However,  the goal here is rather to show  how MAN arises naturally in the context of  information transmission physical scenarios.

\subsection{Basic properties}
From Eq.~(\ref{MAN}) it is clear that MAN has the following properties: 
a)  it is a symmetric function of its arguments i.e.,
$S({\cal A}:{\cal B})=S( {\cal B}:{\cal A});$  b) it vanishes iff ${\cal A}\subset {\cal B}^\prime$ ($\Leftrightarrow {\cal B}\subset {\cal A}^\prime$);
c) it is unitarily invariant,  namely for all unitaries $U$ one has that $S({\cal A}: {\cal B})=S({\cal U}({\cal A}): {\cal U}({\cal B}))$
\footnote{Here, if $\cal X$ is an algebra  ${\cal U}({\cal X}):=\{ {\cal U}(X)\,/\, X\in{\cal X}\}$}.

The next proposition  summarizes the other basic  properties of MAN as a function of pairs $({\cal A}, {\cal B}).$ 
of algebras.
\begin{proposition}\label{prop: prop-1}
Let  $ \{e_\alpha\}$  and $ \{f_\beta\}$ denote the orthogonal bases of $\cal A$  and $\cal B$  defined in Eq.~(\ref{eq:bases}), then.
\begin{itemize}
\item[i)] 
\begin{eqnarray}\label{eq:Omega}
S({\cal A}: {\cal B})&=&1-\frac{1}{d}\mathrm{Tr}\left(S\, \Omega_{\cal A} \Omega_{\cal B}\right) \nonumber\\
 &=&\frac{1}{2d}\sum_{\alpha,\beta}\| [e_\alpha,\, f_\beta]  \|_2^2
\end{eqnarray}
where $\Omega_{\cal A}:=\sum_{\alpha=1}^{d({\cal A})}e_\alpha\otimes e_\alpha^\dagger,$ and $\Omega_{\cal B:}=\sum_{\beta=1}^{d({\cal B})}f_\beta\otimes f_\beta^\dagger$
and $S$ is the swap over ${\cal H}^{\otimes\,2}.$
 \item[ii)] 
If  $\mathbb{P}_{{\cal B}^\prime}$ is the projection CP-map over ${\cal B}^\prime$ 
one has
\begin{align} \label{eq:Q}
 S({\cal A}:{\cal B})= S({\cal A} | {\cal B}^\prime)=1- \frac{1}{d}\sum_{\alpha=1}^{d({\cal A})}\|\mathbb{P}_{{\cal B}^\prime}(e_\alpha) \|_2^2
 \end{align}
 \begin{align}\label{eq:P-S_2}
 S_2({\cal A}:{\cal B})=-\log\left(  \frac{1}{d}\sum_{\alpha=1}^{d({\cal A})} \|   \mathbb{P}_{{\cal B}^\prime} (e_\alpha) \|_2^2     \right).
 \end{align}
 Because of i)  an analogous expression holds with $\cal A$ and $\cal B$ exchanged.
\item[iii)] Monotonicity 
\begin{align}\label{eq:monotone}
 & & {\cal A}_1\subset {\cal A}_2\Rightarrow S({\cal A}_1: {\cal B})\le S({\cal A}_2: {\cal B}),\,(\forall {\cal B})
\end{align}
same holds with respect to $ {\cal B}$
and  for  $S_2$.
\item[iv)] Upper bounds
\begin{align}\label{eq:upper-bound}
S({\cal A}: {\cal B})\le 1- \frac{1}{d^2}\max\{d({\cal A}^\prime), d({\cal B}^\prime)  \}.
\end{align}
\begin{align}\label{eq:upper-bound-2}
S_2({\cal A}: {\cal B})\le \log d^2-  \max \{\log d({\cal A}^\prime),  \log d(\cal B) \}
\end{align}
One has also the weaker bound $S({\cal A}: {\cal B}) \le 1- 1/\min \{d({\cal A}), d({\cal B}) \}$ and a similar one for $S_2.$
\end{itemize}
\end{proposition}

Points i) and ii) above provide  analytically  viable definitions of MAN.  Eqs.~ (\ref{eq:Omega}) and (\ref{eq:Q}) are a rather straightforward  extension of results obtained in  \cite{Zanardi-GAAC-2021, A-OTOC-Faidon-2022} for specific choices of algebras $\cal A$ and $\cal B$. 
In particular ii) shows explicitly that the only basis elements $e_\alpha\in{\cal A}$ that contribute to it (\ref{MAN}) are those which are {\em{not}} in ${\cal B}^\prime.$ The latter  is,  in this sense, ``subtracted" from $\cal A$.  Point iii) shows the key monotonicity property of  MAN: enlarging either $\cal A$ or $\cal B$ will not decrease it.  
The upper bounds iv) are then obtained by  setting $\cal B$ as large as possible i.e., ${\cal B}=L({\cal H}).$
In this case: 
\begin{align}\label{eq:fraction}
S({\cal A}: L({\cal H}))=1-\frac{d({\cal A}^\prime)}{d^2}=\frac{\mathrm{dim}\left( L({\cal H})/{\cal A}^\prime\right)}{\mathrm{dim}\, L({\cal H})}.
\end{align}
Intuitively,  the MAN of a subalgebra with the full operator algebra is given by the ``fraction" of operators outside its commutant.
Colloquially,  one could also say that  (\ref{eq:fraction}) is the probability that an operator chosen uniformly at  random in $L({\cal H}) $  falls outside of  ${\cal A}^\prime.$

{\em{Remark}}. 
By generalizing results for $\cal A$-OTOC's  averaged over unitary channels [see Eq.~(13) in  Ref.~\cite{A-OTOC-Faidon-2022}]
one can average MAN over a unitary orbit of $\cal B$ (or $\cal A$)
\begin{align}\label{eq:averaged-MAN}
\mathbb{E}_U\left[  S({\cal A}:{\cal U}({\cal B}))    \right]=\frac{ S({\cal A}:L({\cal H}))\,S({\cal B}:L({\cal H}))}{ S(L({\cal H}):L({\cal H}))    }.
\end{align}
This can be interpreted as the probability that an operator falls outside both ${\cal A}^\prime$ {\em{and}}  ${\cal B}^\prime,$ conditioned on  not being a scalar multiple of the identity.

\subsection{ MAN and $\cal A$-OTOCs} \label{subsec:A-OTOC} 
A first example,  and a key physical motivation,  of the MAN  formalism is provided by the
$\mathbf{\cal A}$-OTOC  $G_{\cal A}({\cal U})$  introduced in \cite{A-OTOC-Faidon-2022} to study quantum scrambling and chaos at level of full algebras of observables: 
\begin{align}\label{eq:MAN-A-OTOC}
G_{\cal A}({\cal U})=S({\cal A}: {\cal U}({\cal A}^\prime))=S({\cal A}|{\cal U}({\cal A})).
\end{align}
Namely,  the $\cal A$-OTOC is simply the MAN of the pair given by an algebra ${\cal A}$ and the unitary image of its commutant [note that ${\cal U}({\cal A}^\prime)={\cal U}({\cal A})^\prime$].

In particular when:
 {\em{a)}} ${\cal H}={\cal H}_A\otimes{\cal H}_B$ and ${\cal A}=L({\cal H}_A)\otimes\mathbf{1}_B, $  Eq.~(\ref{eq:MAN-A-OTOC}) 
is the averaged bipartite OTOC introduced in \cite{styliaris_information_2020} which, in turn,  coincides with the operator entanglement of $U$ \cite{zanardi2001entanglement}.  
 {\em{b)}} ${\cal A}=\mathbf{C}\{ |i\rangle\langle i|\}_{i=1}^d$ is maximal abelian algebra.,  Eq.~(\ref{eq:MAN-A-OTOC})
is the Coherence Generating Power of $U$ defined in  \cite{zanardiCoherencegeneratingPowerQuantum2017}.

These remarks show that MAN has already found applications to  quantum scrambling,  coherence and chaos. 
The next example will show another application in the context of a simple definition of relative quantumness. 

\subsection{Maximal Abelian Algebras}
Let ${\cal A}_B$ and ${\cal A}_{\tilde{B}}$ two maximal abelian algebras spanned by the resolutions of the identity
$\{\Pi_i:= |i\rangle\langle i|\}_{i=1}^d$ and $\{ \tilde{\Pi}_j:=|\tilde{j}\rangle\langle \tilde{j}|\}_{j=1}^d$ respectively.
Then $\Omega_{{\cal A}_B}=\sum_{i=1}^d |i\rangle\langle i|^{\otimes\,2}$ and similarly for ${\cal A}_{\tilde{B}}.$
From (\ref{eq:Omega}) one gets
\begin{eqnarray}\label{eq:MAN-MASA}
S({\cal A}_B: {\cal A}_{\tilde{B}})
&=&\frac{1}{d}\sum_{i=1}^d S_{lin}(\mathbf{p}_i),\\
S_2({\cal A}_B: {\cal A}_{\tilde{B}})&=&-\log\left( \frac{1}{d}\sum_{i=1}^d \|\mathbf{p}_i\|^2\right),
\end{eqnarray}
where $S_{lin}(\mathbf{p}):=1-\|\mathbf{p}_i\|^2$ is the {\em{linear entropy}} of the probability vectors 
$\mathbf{p}_i:= (  |\langle i|\tilde{j}\rangle|^2)_{j=1}^d.$
 
 If $|\tilde{i}\rangle=U|i\rangle, \,(\forall i)$ for a unitary $U$ then,  as mentioned in the above,   Eq.~(\ref{eq:MAN-MASA}) amounts to the  Coherence Generating Power of $U$ in the basis
 $B=\{|i\rangle\}_{i=1}^d$ \cite{zanardiCoherencegeneratingPowerQuantum2017}.  The latter is proportional to the square of  to the distance between ${\cal A}_B$ and ${\cal A}_{\tilde{B}}$  \cite{zanardiQuantumCoherenceGenerating2018} [see also Eq.~(\ref{eq:MAN-dist})].  
It follows that the MAN of two maximal abelian algebras vanishes iff they coincide \footnote{This can also be immediately seen from the fact that $S({\cal A}_B: {\cal A}_{\tilde{B}})=0$
iff  ${\cal A}_B\subset {\cal A}_{\tilde{B}}^\prime=  {\cal A}_{\tilde{B}}$ and since,  by symmetry of $S$,  even the opposite inclusion holds the  
$ {\cal A}_{\tilde{B}}= {\cal A}_{{B}}$  identity follows. }.

To gain further insight into the physical meaning of MAN in the case of maximal abelian algebras we  define the {\em{quantumness}} of the algebra ${\cal A}_{\tilde{B}}$ {\em{relative}}
to ${\cal A}_B$ by
\begin{eqnarray}\label{eq:Quantumness}
Q_{B}(\tilde{B}):= \sup_{A}\,\{  \frac{1}{d}\sum_{i=1}^d \sigma_i^2(A)\}
\end{eqnarray}
where $\sigma_i^2(A):= \langle i| A^2|i\rangle -\langle i|A|i\rangle^2,$ is the variance of $A$ in the state $|i\rangle\in B$ and the supremum is taken over $A=A^\dagger\in {\cal A}_{\tilde{B}}$ such that $\|A\|_2=1.$ 

Clearly,  (\ref{eq:Quantumness}) vanishes  iff  all the observables $A\in{\cal A}_{\tilde{B}}$ are variance-free  in all the states of the basis $B.$  In this case,  since these  observables have no quantum fluctuations,  they can be called classical relative to $B.$
One can  prove the following bounds 
\begin{align}\label{eq:Quantumness-upper-bound}
Q_{B}(\tilde{B})\le S({\cal A}_B: {\cal A}_{\tilde{B}})\le d\, Q_{B}(\tilde{B}),
\end{align}
which imply $ Q_{B}(\tilde{B})=0\Leftrightarrow S({\cal A}_B: {\cal A}_{\tilde{B}})=0$  \footnote{See Appendix \ref{sec:proof-quantumness-bound}}.

The second inequality above shows that,  as soon as the two algebras differ i.e., MAN is not zero,  there must exist some observable in ${\cal A}_{\tilde{B}}$
for which some of the states $|i\rangle\in B$ feature quantum fluctuations.
Conceptually,  each  algebra ${\cal A}_B$ defines a sort of  maximal classical domain embedded in the Hilbert space and the MAN between a pair of these domains bounds their relative quantumness. The latter vanishes  iff the two classical domains are in fact identical.
 
\subsection{Factors and Abelian algebras}
Let us now consider an example in which the  first algebra is a factor and the second is Abelian.
Let ${\cal H}={\cal H}_1\otimes {\cal H}_2,\,(d_j=\mathrm{dim}({\cal H}_j,) \,j=1,2)$ and $${\cal A}= L({\cal H}_1)\otimes\mathbf{1}_2,\quad{\cal B}=\mathbf{C}\{\Pi_i\}_{i=1}^{d({\cal B})}$$ where the $\Pi_i$ are an orthogonal resolution of the identity.    
Following \cite{Zanardi-GAAC-2021} one finds $$\Omega_{\cal A}=\frac{1}{d_1} S_ {11^\prime},\quad \Omega_{\cal B}=\sum_{i=1}^{d({\cal B})}\Pi_i\otimes\Pi_i,$$ where $S_ {11^\prime}$ is swap of the first factors  only between the two copies in ${\cal H}^{\otimes\,2}=({\cal H}_1\otimes {\cal H}_2)\otimes ({\cal H}_1\otimes {\cal H}_2).$
Therefore,  from (\ref{eq:Omega})  [or  (\ref{eq:Q})]  it follows
\begin{align}\label{eq:example-Abelian-Factor}
S({\cal A}: {\cal B})=1-\frac{1}{d\,d_1}\sum_{i=1}^{d({\cal B})} \|  \mathrm{Tr}_1(\Pi_i) \|_2^2.
\end{align}
Let us now specialize to the case $d({\cal B})=d=d_1 d_2$ i.e., $\cal B$ is maximal and  $\Pi_i=|i\rangle\langle i|,(\forall i).$

The following facts can be easily checked:\\
{{i)}} Eq. ~(\ref{eq:example-Abelian-Factor}) becomes the average (quantum) linear  entropy of the bi-partite states $\rho_i:= \mathbf{1}/d_1\otimes \mathrm{Tr}_1( \Pi_i)=
\mathbb{P}_{{\cal A}^\prime}(\Pi_i)$ i.e.,  
$S({\cal A}: {\cal B})=\frac{1}{d} \sum_{i=1}^d (1-\|\rho_i\|_2^2).$\\
   {{ii)}}  
   The upper bound (\ref{eq:upper-bound}) is achieved when all of the $|i\rangle$'s are maximally entangled i.e.,  $\| \mathrm{Tr}_1( \Pi_i) \|_2^{2}=1/\min\{d_1, \, d_2\} \,(\forall i).$ \\
  {{iii)}}  Eq. ~(\ref{eq:example-Abelian-Factor})   is lower bounded by $1-1/d_1$ which is achieved when all of the $|i\rangle$'s  are product states.\\
   {{iv)}} The log-MAN is given by  
  \begin{align}\label{eq:example-max-Abelian-Factor-2}
S_2({\cal A}: {\cal B})= -\log\left(\frac{1}{d}\sum_{i=1}^d \|\rho_i\|_2^2 \right)\le \frac{1}{d}\sum_{i=1}^d S_2(\rho_i)
\end{align}
where $ S_2(\rho):= -\log \|\rho\|_2^2$ is the 2-Renyi entropy.  The upper bound is achieved as in ii). 

The results above for the maximal abelian $\cal B,$ make it clear that one can construct a MAN bearing a direct relation to the entaglement content of a whole orthonormal system. 

Moreover,  they suggest that {\em{MAN is morally a quantity of entropic nature}}.  
This is consistent with the fact that the $\cal A$-OTOC (i.e.,  a specific type of MAN) discussed in  \cite{Zanardi-GAAC-2021, A-OTOC-Faidon-2022} have a similar interpretation e.g.,  operator entanglement entropy \cite{zanardi2001entanglement}  for factors,  and entropy of probability vectors for maximal abelian algebras as seen in Eq.~(\ref{eq:MAN-MASA}). 
In Sect.~(\ref{sec:morally}) we show how, even in the most general case, MAN can be seen as an average (linear) entropy production  of a family of  CP maps associated to $\cal A$ and $\cal B$

\subsection{The net of local algebras}
In order to further  illustrate the MAN formalism developed in this paper we will now consider the net of local algebras associated with a quantum lattice system.

Let $\Lambda$ be a set  with finite cardinality $|\Lambda|$ such that for each element $i\in\Lambda$ there is an associated Hilbert space ${\cal H}_i\cong \mathbf{C}^d$.
For any subset $S\subset \Lambda$ one can build the tensor product ${\cal H}_S:=\otimes_{i\in S} {\cal H}_i\cong  \mathbf{C}^{d^{|S|}}.$ In this way one is given the following family of factor algebras
\begin{align}
\mathbb{A}_{loc}:=\{S\subset\Lambda\,/\, {\cal A}_S:=L({\cal H}_S)\otimes\mathbf{1}_{S^c}\}.
\end{align}
Here $S^c$ is the  complement of $S$, and ${\cal A}_\emptyset=\mathbf{C}\mathbf{1}$ \footnote{
The following facts hold true: 
i) $d({\cal A}_S)=(d^2)^{|S|}$.  ii) $S_1\subset S_2\Rightarrow {\cal A}_{S_1}\subset  {\cal A}_{S_2}$.  iii) $ {\cal A}_{S}^\prime= {\cal A}_{S^c}$.  iv) ${\cal A}_{S_1}\cap 
{\cal A}_{S_2}={\cal A}_{S_1\cap S_2}$.  v) ${\cal A}_{S_1}\vee 
{\cal A}_{S_2}={\cal A}_{S_1\cup S_2}$. vi) $\mathbb{P}_{{\cal A}_{S_1}}\mathbb{P}_{{\cal A}_{S_2}}=\mathbb{P}_{{\cal A}_{S_2}}\mathbb{P}_{{\cal A}_{S_1}}=\mathbb{P}_{{\cal A}_{S_1\cap S_2}}$.
}.

\begin{proposition}

\begin{itemize}
 \item[i)] If $c_d:=  \log d^2.$
\begin{align} \label{eq:S-loc}
 S({\cal A}_{S_1}: {\cal A}_{S_2})=1-\frac{1}{d^{2\,|S_1\cap S_2|}},
 \end{align}
\begin{align}\label{eq:S_2-loc}
S_2({\cal A}_{S_1}: {\cal A}_{S_2})= c_d\,|S_1\cap S_2| 
\end{align}
\item[ii)] 
\begin{align}
\label{eq:NC-loc}
NC({\cal A}_{S})=1-\frac{1}{d^{2\,|S|}},\quad NC_2({\cal A}_S)= c_d |S|
\end{align}
\end{itemize}
\end{proposition}
These results show the simple geometrical meaning of the log-MAN for the net of local algebras: it is the ``volume" of the joint spatial support of the two algebras.  
Notice that from the above it also follows $S_2({\cal A}_{S_1}|{\cal A}_{S_2})=c_s|S_1 \cap S_2^c |=c_d |S_1 \setminus  S_2|,$ which further illustrates the geometrical meaning of the MANs (and the adopted notation).

These remarks seem to suggest a ``non-commutative spatial"  interpretation of the log-MANs,  in the general case,  as `` volumes" of the would-be  algebra supports.  
Abelian algebras have zero non-commutative (or ``quantum volume") and factors a maximal one.
\section{Self-MAN}
%
In this section we will focus on the self-MAN $NC({\cal A})=S({\cal A}:{\cal A}).$ From unitary invariance of MAN one has  
that the self-MAN is constant along the equivalence class of unitarily isomorphic algebras.
Le us now suppose that the algebra $\cal A$ has the following  central-blocks decomposition  [See Eq.~(\ref{eq:hilb-decom})]
\begin{align}\label{eq:alg-decomp}
{\cal A}\cong \bigoplus_{J=1}^{d_Z}  \mathbf{1}_{n_J}\otimes L( \mathbf{C}^{d_J}).
\end{align}
The next result shows that $NC$ depends just on the structural  data $\mathbf{d}=(d_J)_J$ and  $\mathbf{n}=(n_J)_J.$
An algebra is called {\em{collinear}} is the vectors $\mathbf{d}$ and $\mathbf{n}$ are collinear  i.e.,  $n_J/d_J$ is $J$-independent.  Factors and maximal abelian algebras are clearly examples of collinear algebras.
 \begin{proposition}
 \begin{itemize}
 \item[i)]  
 \begin{align}\label{eq:NC}
 NC({\cal A})=1-\frac{1}{d}\sum_J^{d_Z} \frac{n_J}{d_J} \le 1-\frac{1}{\max_J d_J^2} 
 \end{align}
  \item[ii)]
  \begin{align}\label{eq:NC-upper-bound}
  NC({\cal A})\le 1-\frac{1}{d^2}=NC(L({\cal H}))
  \end{align} 
 \item[i)] If ${\cal A}$ is collinear
 \begin{align}\label{eq:NC-coll}
 NC({\cal A})=1-\frac{d_Z}{d({\cal A})}=\frac{d({\cal A}/{\cal Z}({\cal A}))}{d({\cal A})}
 \end{align}
 here ${\cal Z}({\cal A})):={\cal A}\cap{\cal A}^\prime$ is the center of $\cal A$ whose dimension is $d_Z$.  In particular, if $\cal A$ is a factor ($d_Z=1$)
 the upper bound (\ref{eq:NC}) is saturated.
 \end{itemize}
 \end{proposition}
 Point i) above provides an explicit,  and quite simple,  formula for the self-MAN in terms of the data of the central-blocks decomposition (\ref{eq:alg-decomp}).
 Point ii) shows that the maximal value of the self-MAN is achieved by the full $L({\cal H})$.  Finally,  point iii) shows that in the collinear case
 the self-MAN is the relative dimension of the algebra after its center has been modded out.  More loosely speaking,  the self-MAN is given by the fraction of elements of the algebra that are not inside the commutant as well [see similarity with Eq. ~(\ref{eq:fraction})].
 
{\em{Remarks.}}

{\em{a)}}
Eq.~(\ref{eq:NC}) can written as an average of  $NC$'s of factors $NC({\cal A})=\sum_J^{d_Z} p_J\left(1-\frac{1}{d_J^2} \right)=\sum_J p_J NC ({\cal A}_{d_J}),$ where
 $p_J:=\frac{n_J d_J}{d},$ is a $Z$-dimensional probability and ${\cal A}_{d_J}:= \mathbf{1}_{n_J}\otimes L(\mathbf{C}^{d_J})$
 are factors over the algebra irreps $\mathbf{C}^{d_J}.$ In words: the self-MAN of the algebra is the mean of the self-MAN's of its irreps
 weighted by the fractional dimension of the corresponding central block.
 
{\em{b)}} The log-self-MAN for a general algebra $\cal A$ can be cast in a form similar to (\ref{eq:example-max-Abelian-Factor-2}):
\begin{align}\label{eq:log-Self-MAN}
NC_2({\cal A})=-\log\left(   \sum_J p_J \| R_J  \|_2^2       \right),
\end{align}
where  $R_J:=(\frac{\mathbf{1}_{d_J}}{d_J})^{\otimes 2}$
are doubled maximally mixed states.
From convexity,  $NC_2({\cal A})\le \sum_J p_J S_2(R_J)\le \max_J S_2(R_J)=\log (\max_J d_J^2).$  
Again,  these bounds are saturated by factors and point,  one more time,  to the entropy-like character of $NC.$

Let us now discuss two examples with small and large self-MAN respectively.

 \subsection{Asymptotically Abelian}
 Consider the following Hilbert space decomposition
 \begin{align}\label{eq:almost-abelian}
 {\cal H}=\left(\oplus_{i=1}^{d-2}\mathbf{C}\otimes \mathbf{C}\right) \oplus \mathbf{C}\otimes \mathbf{C}^2,
 \end{align}
 and the corresponding algebra ${\cal A} =\left(\oplus_{i=1}^{d-2}\mathbf{1}_1\otimes \mathbf{1}_1\right) \oplus \mathbf{1}_1\otimes L(\mathbf{C}^2).$
 This is a $(d-1)-$dimensional maximal abelian algebra glued with one non-commutative qubit block.  Note that $d({\cal A})=d-2+2^2=d+2,$ and $d_Z=d-1.$
 Using (\ref{eq:NC}) one finds 
\begin{align} 
 NC({\cal A})=1-\frac{1}{d}\left(  (d-2)\frac{1}{1} +\frac{1}{2}  \right)=\frac{3}{2d}
 \end{align}
 For large $d$ this algebra is asymptotically abelian in the sense that $NC({\cal A})=O(1/d).$ 
 \subsection{Symmetric operators}
 Let ${\cal H}=(\mathbf{C}^d)^{\otimes\,2}$ and ${\cal A}=\mathbf{C}\{\mathbf{1},\, S\}^\prime$ where $S$ is the swap operator i.e., ${\cal A}$ is the algebra of symmetric operators.
   The decomposition (\ref{eq:hilb-decom}) now reads ${\cal H}=\sum_{J=\pm} \mathbf{C}\otimes\mathbf{C}^{d_J},$
 where $d_{\pm}=d(d\pm 1)/2.$ Since $n_\pm=1$ Eq.~(\ref{eq:NC}) is given by
\begin{align}\label{eq:self-MAN-example}
 NC({\cal A})=1-\frac{1}{d^2}\sum_{J=\pm}\frac{1}{d_J}=1- \frac{4}{d^2(d^2-1)}.
\end{align} 
  Notice that 
  $d({\cal A})=d_+^2+d_-^2= d^2(d^2+1)/2,$  and $d_Z=2.$  It follows that   
 the expression (\ref{eq:NC-coll}) is a {\em{strict}} upper bound for (\ref{eq:self-MAN-example}) at any finite $d$\footnote{This is true for any $\cal A$ with abelian commutant i.e., $n_J=1\,\forall J$}. Moreover,  since $d({\cal A}^\prime)=2$ also
 the upper bound  (\ref{eq:upper-bound}) (with ${\cal B}={\cal A}$) is always strict.  These facts  are consequences of  $\cal A$ being   {\em{not}} collinear (See also next Proposition).
 This is a highly non-commutative example as  (\ref{eq:self-MAN-example}) is $O(1/\mathrm{dim}^2({\cal H}))$ away from the absolute upper-bound (\ref{eq:NC-upper-bound})\footnote{Note that here $\mathrm{dim} {\cal H}=d^2$ thus (\ref{eq:NC-upper-bound}) reads $\le 1-1/d^4$}.
 \section{MAN,  projections and protocols}\label{sec:alg-proj}
 In this section we will focus on the case in which at least one of the two algebras, say $\cal A,$ in the definition (\ref{MAN}) is {\em{collinear}}.
Indeed,  in this case  MAN {\em{can be expressed in terms of the algebra projection superoperators}} (See Appendix.).  Since the latter are unital CP-maps they correspond to quantum physical processes which are in principle realizable.
  \begin{proposition}
 If $\cal A$ is collinear then
 \begin{itemize}
 \item[i)]
 \begin{align}\label{eq:MAN-Proj}
 S({\cal A}: {\cal B})=1-\frac{1}{d({\cal A})} \mathrm{Tr}_{HS}\left( \mathbb{P}_{\cal A}\, \mathbb{P}_{{\cal B}^\prime}\right).
 \end{align}
 \item[ii)] If $d({\cal A})=d({\cal B}^\prime)$ then 
\begin{align}\label{eq:MAN-dist} 
 S({\cal A}: {\cal B})=\frac{1}{2\,d({\cal A})} \| \mathbb{P}_{\cal A} -\mathbb{P}_{{\cal B}^\prime}   \|_{HS}^2.
 \end{align}
 \item[iii)]
 \begin{eqnarray}\label{eq:upper-bound-intersec}
 S({\cal A}: {\cal B})&\le& 1-\frac{d({\cal A}\cap {\cal B}^\prime)}{d({\cal A})}.
\end{eqnarray}
 This bound is saturated when $[\mathbb{P}_{\cal A},\,\mathbb{P}_{{\cal B}^\prime}]=0$.
 \end{itemize}
 \end{proposition}
Point ii) shows that when ${\cal B}^\prime$ has the same dimension of $\cal A$ then MAN has a simple geometrical interpretation: it is proportional to the square of the distance $D({\cal A}, {\cal B}^\prime):=  \| \mathbb{P}_{\cal A} -\mathbb{P}_{{\cal B}^\prime}   \|_{HS}$ between $\cal A$ and ${\cal B}^\prime.$ This is the situation relevant to the case of the $\cal A$-OTOC introduced in \cite{A-OTOC-Faidon-2022} to study quantum scrambling.  Point iii) establishes an upper bound neatly involving the  dimension of the intersection algebra ${\cal A}\cap {\cal B}^\prime\supset\mathbf{C}\mathbf{1}.$
Colloquially,   one could  say that the MAN between $\cal A$ (collinear) and $\cal B$ is upper bounded  by the ``probability" that an element of $\cal A$ lies outside of the commutant of $\cal B$.
Note,  Eq.~(\ref{eq:fraction}) and (\ref{eq:NC-coll}) are special cases  where  (\ref{eq:upper-bound-intersec})  is saturated. 

If one assumes the  {\em{ability to implement}} the projection CP-maps as physical processes one can devise  protocols to measure the MAN in the laboratory.  
The next two propositions provide the mathematical ground for two of such operational interpretations.  Basically,  the point is that MAN will be expressed in terms of expectation values of swaps over states that can be prepared by the projections CP-maps.
\subsection{Algebra States}
We first describe  a protocol for MAN's estimation  based on ``algebra states" over a doubled Hilbert space.  In  the self-MAN case the problem is reduced to the estimation of the purities of these algebra states.

 Using the standard Choi–Jamiołkowski  construction and the projection CP-maps one can associated quantum states over ${\cal H}^{\otimes\,2}$ to algebras
\begin{align}\label{eq:choi}
\omega({\cal A}):=(\mathbb{P}_{\cal A}\otimes\mathbf{1})(|\Phi^+\rangle\langle\Phi^+|),
\end{align}
where $|\Phi^+\rangle:=\frac{1}{\sqrt{d}}\sum_{i=1}^d |i\rangle\otimes|i\rangle.$ 

\begin{proposition}
\begin{itemize}
\item[i)] If $S$ is the swap over $({\cal H}^{\otimes\,2})^{\otimes\,2}$
\begin{align}
\label{eq:MAN-choi}
S({\cal A}: {\cal B})=1-\frac{\mathrm{Tr}\left( S\, \omega({\cal A})\otimes \omega({\cal B}^\prime)\right)}{\|\omega({\cal A})\|_2^2}
\end{align}
\item[ii)] In particular for ${\cal A}={\cal B}$
\begin{align}\label{eq:self-MAN-choi}
NC({\cal A})=1-\frac{ \|\omega({\cal Z}({\cal A}))\|_2^2 }{\|\omega({\cal A})\|_2^2},
\end{align}
equivalently 
\begin{align}\label{eq:2-self-MAN-choi}
NC_2( {\cal A})=S_2(\omega({\cal Z}({\cal A}))-S_2(\omega({\cal A})),
\end{align}
where $S_2(\omega):=-\log\|\omega\|_2^2$ is the $2$-Renyi entropy of the state $\omega.$
\end{itemize}
\end{proposition}
 Eq.~(\ref{eq:2-self-MAN-choi})  shows in yet another way that  self-MAN  is an entropy-like quantity:
$NC_2({\cal A})$ can be seen as the additional entropy generated by the channel $\mathbb{P}_{{\cal A}^\prime}\otimes\mathbf{1}$ acting on the algebra state $\omega({\cal A})$ \footnote{$\omega({\cal Z}({\cal A}))=(\mathbb{P}_{{\cal Z}({\cal A})}\otimes\mathbf{1})|\Phi^+\rangle\langle\Phi^+|=
(\mathbb{P}_{{\cal A}^\prime}\otimes\mathbf{1}))( \mathbb{P}_{{\cal A}}\otimes\mathbf{1}) |\Phi^+\rangle\langle\Phi^+|=(\mathbb{P}_{{\cal A}^\prime}\otimes\mathbf{1}) \omega({\cal A})
$}. 

{\bf{Operational protocol 1}} 

To estimate the numerator in (\ref{eq:MAN-choi})

{\bf{1)}}  Prepare two copies of the singlet $|\Phi^+\rangle\in{\cal H}^{\otimes\,2}.$\\
{\bf{2)}} Process the first (second) copy with the CP-map $\mathbb{P}_{\cal A}\otimes\mathbf{1}$ ($\mathbb{P}_{{\cal B}^\prime}\otimes\mathbf{1}$). \\
{\bf{3)}} Measure the expectation of $S$   (Swap over $({\cal H}^{\otimes 2} )^{\otimes 2} $).

To estimate the denominator  in (\ref{eq:MAN-choi}): 
Steps {\bf{1)}} and {\bf{3)}} are the same.  In step  {\bf{2})}  apply  $\mathbb{P}_{\cal A}\otimes\mathbf{1}$ to both copies.
Note,  this also provides an estimate of the relative algebra dimension as $\|\omega({\cal A}) \|_2^2= \mathrm{Tr}\left[ S \omega({\cal A})^{\otimes\,2}\right] =d({\cal A})/d^2.$

\subsection{Stochastic Protocol}
We now describe a stochastic protocol for MAN's estimation.  In  the self-MAN case the problem is reduced to the estimation of the purities of the states obtained by the action of the algebra projections on Haar random states in $\cal H$.
\begin{proposition}
\begin{itemize}
 \item[i)]
\begin{align}\label{eq:MAN-op}
S({\cal A}: {\cal B})=1-\frac{ \mathbb{E}_\phi\left[  \langle  \mathbb{P}_{\cal A}(\hat{\phi}),\,    \mathbb{P}_{{\cal B}^\prime}(\hat{\phi})\rangle  \right]-\frac{1}{d+1}}
{ \mathbb{E}_\phi\left[  \|\mathbb{P}_{\cal A}(\hat{\phi})\|_2^2  \right] -\frac{1}{d+1}}.
\end{align}
\item[ii)] In particular for ${\cal A}={\cal B}$
\begin{align}\label{eq:self-MAN-op}
NC({\cal A})=1-\frac{ \mathbb{E}_\phi\left[  \|\mathbb{P}_{{\cal Z}({\cal A})}(\hat{\phi})\|_2^2  \right]-\frac{1}{d+1}}
{ \mathbb{E}_\phi\left[  \|\mathbb{P}_{\cal A}(\hat{\phi})\|_2^2  \right] -\frac{1}{d+1}}.
\end{align}
\end{itemize}
In the above $\hat{\phi}:=|\phi\rangle\langle\phi|$ is a one-dimensional projector and the expectations are with respect to the Haar induced distribution of the $|\phi\rangle.$
\end{proposition}

{\bf{Operational protocol 2}} 

To estimate the numerator in (\ref{eq:MAN-op})
 
{\bf{1)}}  Prepare two copies of the Haar random state $|\phi\rangle$.\\
{\bf{2)}} Process the first (second) copy with the CP-map $\mathbb{P}_{\cal A}$ ($\mathbb{P}_{{\cal B}^\prime}$).\\
{\bf{3)}} Measure the expectation of $S$ (Swap over ${\cal H} \otimes{\cal H}$).\\
{\bf{4)}} Repeat steps {\bf{1)}}--{\bf{3)}} and average results.

To estimate the denominator  in (\ref{eq:MAN-op}): 
Steps {\bf{1)}} and {\bf{3),4)}} are the same.  In step  {\bf{2)}}  apply  $\mathbb{P}_{\cal A}$ to both copies.

{\em{Remark.--}} The averaged functions in the equations above are Lipschitzian \footnote{
For example,   if $f(\phi):=  \|\mathbb{P}_{\cal A}(\hat{\phi})\|_2^2=\mathrm{Tr}\left[
S\, \mathbb{P}_{\cal A}^{\otimes 2}  (\hat{\phi}^{\otimes 2})\right]$ then, using standard operator norm inequalities and the fact that the algebra projections are CP-maps,  $|f(\phi_1)-f(\phi_2)|\le \|  \mathbb{P}_{\cal A}^{\otimes 2}  (\hat{\phi}_1^{\otimes 2} -\hat{\phi}_2^{\otimes 2})\|_1\le \|\hat{\phi}_1^{\otimes 2} -\hat{\phi}_2^{\otimes 2}\|_1
\le K \| |\phi_1\rangle- |\phi_2\rangle\|$ where $K=O(1)$} and therefore the Levy Lemma for measure concentration applies.  This means,  that in large dimension $d,$ typically very few input $|\phi\rangle$'s will be required for a reliable empirical estimate of the expectations in Eqs.~(\ref{eq:MAN-op}) and  (\ref{eq:self-MAN-op}).

\section{Conclusions}\label{sec:conclusions}
In this paper we introduced an elementary  quantitative  measure of non-commutativity between two algebras of quantum operators acting on common Hilbert-Space.

This quantity,  which we called  Mutual Averaged Non-commutativity (MAN), is defined by a Haar averaged squared norm of the  commutator  between unitaries belonging to the two algebras.   

Specific instances  of MAN have already found physical applications.  Indeed,  the $\cal A$-OTOC recently introduced in the study of quantum scrambling and chaos  \citep{A-OTOC-Faidon-2022}, is recovered when  the second algebra is the unitary image of the commutant of the first.

 Properties and bounds of  MAN have been established in general and we have  shown how MAN   probabilistically   limits the ability to unitarily trasmit information between the two algebras.   In the case of two maximal abelian algebras the MAN provides a natural bound on their ``relative quantumness" and has an elegant metric interpretation.  
 
 We discussed a few significant examples and,  based on the case of the local factors over a lattice system,  suggested that the logarithmic version of MAN 
 can be interpreted as a ``non-commutative volume" of the would-be spatial support of the algebra.  Examples also show that MAN is an entropy-like quantity \footnote{In Sect. ~(\ref{sec:morally}) it is shown how MAN,  {\em{in general}}, can be connected to average entropies of maps associated to $\cal A$ and $\cal B$.}.
 
  When the two algebras coincide the corresponding self-MAN can be fully computed in terms of the dimensions associated with the Hilbert space decomposition induced by the algebra.  
This is a measure of how the algebra is far from being Abelian.  For the physically important class of collinear algebras self-MAN  can be intuitively thought of as the probability that an operator drawn uniformly at random from the algebra falls outside the center.   
 
Furthermore,  for this class of algebras we have shown  how MAN can be expressed in the terms of the algebra projection CP-maps.  Assuming that the latter can be enacted as physical processes, we have described two  operational  protocols which  could be used to to directly estimate the MAN in the laboratory.   

The first protocol is based on ``algebra states" in the doubled Hilbert space and the second on random states processed by the algebra projections.  In the self-MAN case both protocols involve just the measurement of the purity of quantum states.

Unveiling further the operational meaning of the MAN and computing  it explicitly for a wider class of physically motivated algebras  
 constitute   natural goals  for  future investigations.

{\section{Acknowledgments}}
  The author acknowledges discussions with Faidon Andreadakis,  Emanuel Dallas,  Alioscia Hamma  and partial support from the NSF award PHY-2310227.  He is also thankful for hospitality to the Physics Department of the Indian Institute of Technology Bombay where the final part of this work has been carried out.
 This research was (partially) sponsored by the Army Research Office and was accomplished under Grant Number W911NF-20-1-0075.  The views and conclusions contained in this document are those of the authors and should not be interpreted as representing the official policies, either expressed or implied, of the Army Research Office or the U.S. Government. The U.S. Government is authorized to reproduce and distribute reprints for Government purposes notwithstanding any copyright notation herein.

\bibliographystyle{apsrev4-1}
\bibliography{refs, my_library}
\appendix
\begin{widetext}
\section{Basic facts and notation about algebras}\label{sec:basic}

The main ingredients  of this paper are provided by   \emph{hermitian-closed  unital  subalgebras} ${\cal A}\subset L({\cal H})$ and their {\em{commutants}}
\begin{align}\label{eq:commutant}
{\cal A}^\prime:=\{X\in L({\cal H})\,|\, [X,\,Y]=0,\,\forall\, Y\in {\cal A}\}
\end{align}
and {\em{centers}} 
${\cal Z}({\cal A}):={\cal A}\cap {\cal A}^\prime. $
The {\em{double commutant}} theorem holds: $({\cal A}^\prime)^\prime={\cal A}.$
If $\cal A$ (${\cal A}^\prime$) is abelian,  one has that ${\cal A}\subset {\cal A}^\prime$ (${\cal A}^\prime\subset {\cal A}$)
and therefore ${\cal Z}({\cal A})={\cal A}$ (${\cal Z}({\cal A})={\cal A}^\prime$).
A key structural fact used in this paper is that the Hilbert
space splits into a sum of $d_Z:=\mathrm{dim} \,{\cal Z}({\cal A})$  $\cal A$-invariant blocks  and that each of these blocks is  bipartite: 
\begin{align}\label{eq:hilb-decom}
{\cal H}=\bigoplus_{J=1}^{d_Z}  {\cal H}_J,\qquad {\cal H}_J\cong \mathbf{C}^{n_J}\otimes  \mathbf{C}^{d_J},\quad  (d=\sum_J n_Jd_J).
\end{align}
Furthermore,  $\cal A$ (${\cal A}^\prime$) acts irreducibly on the $  \mathbf{C}^{d_J}$ factors ($\mathbf{C}^{n_J}$)
$${\cal A}\cong \bigoplus_{J=1}^{d_Z}  \mathbf{1}_{n_J}\otimes L( \mathbf{C}^{d_J}),\quad{\cal A}^\prime\cong \bigoplus_{J=1}^{d_Z}  L( \mathbf{C}^{n_J})\otimes \mathbf{1}_{d_J}.
$$
 This implies
 $\mathrm{dim}\,{\cal A}=\sum_Jd_J^2=:d({\cal A}),\quad\mathrm{dim}\,{\cal A}^\prime=\sum_Jn_J^2=:d({\cal A}^\prime).\nonumber
 $
 Notice,  $\cal A$ (${\cal A}^\prime$) is {{abelian}} {\em{iff}} $d_J=1,\,(\forall J)$  ($n_J=1,\,(\forall J)$).
 Also,  from the above it follows that the center of $\cal A$ is generated by the projections over the central blocks
 $${\cal Z}({\cal A})=\bigoplus_{J=1}^{d_Z}  \mathbf{C}\{\Pi_J:=\mathbf{1}_{n_J}\otimes \mathbf{1}_{d_J}\}.$$
 %
 
Introducing,  the $d_Z$-dimensional (integer-valued) vectors ${\bf{d}}:=(d_J)_J$, and ${\bf{n}}:=(n_J)_J$,  one has that
 $d^2=(\mathbf{n}\cdot\mathbf{d})^2\le \|\mathbf{n}\|^2\|\mathbf{d}\|^2=
 d({\cal A})d({\cal A}^\prime).$
If  ${\bf{d}}=\lambda {\bf{n}}$, the above inequality becomes an equality, i.e. $d^2=d({\cal A})d({\cal A}^\prime),$
 In this case the pair $({\cal A},\,{\cal A}^\prime)$  (or by abuse $\cal A$ itself) is referred to as  \emph{collinear}.

Affiliated to any algebra $\cal A$, we have an orthogonal projection Completely Positive (CP) map: $\mathbb{P}_{\cal A}^\dagger=\mathbb{P}_{\cal A},\, \mathbb{P}_{\cal A}^2=\mathbb{P}_{\cal A}$ and
$\mathrm{Im}\,\mathbb{P}_{\cal A}={\cal A}.$
 %
  If $U$ is a unitary,  and  ${\cal U}(X):=U XU^\dagger$ and ${\cal U}({\cal A}):=\{ {\cal U}(a)\,|\, a\in{\cal A}\}$ is the image algebra.  The corresponding projection CP-map is   given by
 $\mathbb{P}_{{\cal U}({\cal A})}={\cal U} \mathbb{P}_{{\cal A}} {\cal U}^\dagger.$
 
Corresponding to the structural theorem (\ref{eq:hilb-decom}) one can build  orthogonal bases for $\cal A$ and ${\cal A}^\prime$
 \begin{align}\label{eq:bases}
 e_\alpha:=   \frac{\mathbf{1}_{n_J}}{\sqrt{d_J}}\otimes |l\rangle\langle m|,\quad \tilde{e}_\beta= |p\rangle\langle q|\otimes \frac{\mathbf{1}_{d_J}}{\sqrt{n_J}}
\qquad( \|e_\alpha\|_2^2=\frac{n_J}{d_J},\,\, \|\tilde{e}_\beta\|_2^2=\frac{d_J}{n_J}.)
\end{align} 
 where $J=1,\ldots,d_Z;\; \alpha:=(J,l,m)\;\beta=(J,p,q),\,l,m=1,\ldots,d_J;\; p,q=1,\ldots,  n_J.$
 
In terms of  these bases,  if $X_J:=\Pi_J X \Pi_J,$  one finds the following representation of the Algebra projections  \cite{Zanardi-GAAC-2021}
\begin{align}\label{eq:proj-basis}
\mathbb{P}_{\cal A}(X)=
\sum_\beta \tilde{e}_\beta X \tilde{e}_\beta^\dagger=\sum_J \frac{\mathbf{1}_{n_J}}{{n_J}}\otimes \mathrm{Tr}_{n_J}(X_J),
\qquad
\mathbb{P}_{{\cal A}^\prime}(X)=\sum_\alpha e_\alpha X e_\alpha^\dagger=\sum_J \mathrm{Tr}_{d_J}(X_J)\otimes \frac{\mathbf{1}_{d_J}}{{d_J}}
\end{align}

\section{Hilbert-Schmidt scalar product of maps}
Let ${\cal H}$
be a $d$-dimensional Hilbert space and $L({\cal H}) $ its full operator algebra. 
$L({\cal H}) $  has  a Hilbert space structure via  the Hilbert-Schmidt scalar product: $\langle X,\,Y\rangle:=\mathrm{Tr}\left(X^\dagger Y\right)$ and norm
$\|X\|_2^2:=\langle X,\, X\rangle.$
This equips the space of superoperators i.e., $L(L({\cal H}))$ with the scalar product and the corresponding  norm 
\begin{eqnarray}  \label{eq:HS-scalar-product}
\langle {\cal T},\,{\cal F}\rangle_{HS}:=\mathrm{Tr}_{HS}\left({\cal T}^\dagger {\cal F} \right)=
\sum_{l,m} \langle m| {\cal T}^\dagger {\cal F}(|m\rangle\langle l|)|l\rangle
= \sum_{l,m} \langle {\cal T}(|m\rangle\langle l|),\, {\cal F}(|m\rangle\langle l|)\rangle
\end{eqnarray}
\begin{align}
\|  {\cal T}\|_{HS}^2= \langle {\cal T},\,{\cal T}\rangle=\sum_{l,m} \|{\cal T}(|m\rangle\langle l|)\|_2^2.
\end{align}
For example, if ${\cal T}(X)=\sum_i A_i XA_i^\dagger,$ then $\|{\cal T}\|_{HS}^2=\sum_{i,j} |\mathrm{Tr}(A_i^\dagger A_j)|^2.$
Moreover, if $\mathbb{P}$ is an orthogonal  projection, its rank is given by  $\|\mathbb{P}\|_{HS}^2=\mathrm{Tr}_{HS} \mathbb{P}.$

The swap operator $S$ over ${\cal H}^{\otimes\,2}$ is given by  $S=\sum_{lm}|l m\rangle\langle ml|.$
Using  the identity 
\begin{align}\label{eq:swap-lin}
\mathrm{Tr}\left(AB\right)=\mathrm{Tr}\left[ S (A\otimes B)   \right]
\end{align}
one can write (\ref{eq:HS-scalar-product}) as
\begin{eqnarray}\label{eq:HS-scalar-product-S}
\langle {\cal T},\,{\cal F}\rangle_{HS} 
=\sum_{l m}  \mathrm{Tr}\left[ S |l\rangle\langle m|\otimes {\cal T}^\dagger{\cal F}(|m\rangle\langle l|)\right] =  \mathrm{Tr}\left[ S(\mathbf{1}\otimes    {\cal T}^\dagger{\cal F})\sum_{lm} |lm\rangle\langle m l|        \right]
=\mathrm{Tr}\left[ S(\mathbf{1}\otimes    {\cal T}^\dagger{\cal F})(S)\right].
\end{eqnarray}

\section{Proof of Prop.1}
Let us begin by upper-bounding the distance function (\ref{eq:d_B}) where $U$ and $V$ are unitaries in $\cal A$ and  $X$ an element of $\cal B$ such that $1=\|X\|_\infty\le \|X\|_2\le \sqrt{d}.$
\begin{eqnarray}
|\mathrm{Tr}\left[  X ({\cal U}^\dagger(\rho_0)-{\cal V}^\dagger(\rho_0))    \right]|=|\mathrm{Tr}\left[ \rho_0( {\cal U}(X)  - {\cal V}(X)     )\right]|
\le & \|  {\cal U}(X)  - {\cal V}(X)      \|  \le  \|[U,  X]\|_2 + \|[V,   X]\|_2.\nonumber
\end{eqnarray}
%
Now,   $  \|[U,  X]\|_2=   \sqrt{2}  \sqrt{\|X\|_2^2-\mathrm{Re}\langle {\cal U}({X}),  {X}\rangle}, $
 performing the Haar average with respect to  $U$ and $V$, using concavity  and  $\int_{U({\cal A})} dU \,{\cal U}=\mathbb{P}_{{\cal A}^\prime} $ one gets
$$ \mathbb{E}_{U,V}\left[ d_{{\cal B},\rho_0}(U, V)\right]\le 2\sqrt{2 }\,\sqrt{\|X\|_2^2-\|\mathbb{P}_{{\cal A}^\prime}({X})\|_2^2}.$$
By expanding $X=\sum_\beta b_\beta f_\beta$ on the ortho basis $\{f_\beta\}$ of $\cal B$,  the argument of the last square root reads
$
\sum_{\beta, \gamma} \bar{b}_\beta b_\gamma C_{\beta\gamma}.$ The  positive semi-definite matrix $C$ is given by
$
 C_{\beta\gamma}= \delta_{\beta\gamma} \|f_\beta\|_2^2 -\langle \mathbb{P}_{{\cal A}^\prime}(f_\beta),   \mathbb{P}_{{\cal A}^\prime}(f_\gamma) \rangle.
$\\
Using Eq.~(\ref{eq:Q}) and $\sum_\beta\|e_\beta\|_2^2=1$  one can then write
\begin{eqnarray}
\sum_{\beta, \gamma} \bar{b}_\beta b_\gamma C_{\beta\gamma}\le \|b\|^2 \|C\|_\infty \le  \|b\|^2 \|C\|_1=\|b\|^2 \mathrm{Tr}(C)=
 \|b\|^2 d\left(1-\frac{1}{d}\sum_\beta \|  \mathbb{P}_{{\cal A}^\prime}(f_\beta) \|_2^2\right)=\|b\|^2 d \, S({\cal A}:{\cal B}).
 \nonumber
\end{eqnarray}
From $\|X\|_2\le \sqrt{d},$ one gets $\|X\|_2^2=\sum_\beta |b_\beta|^2 \|f_\beta\|_2^2\le d,$ 
whence $|b_\beta|^2\le d /\| f_\beta\|^2_2,\,(\forall\beta).$ Moreover,  $$\|b\|^2\le d({\cal B})  \max_\beta |b_\beta|^2\le d({\cal B}) d  \max_\beta 1 /\| f_\beta\|^2_2 =:d({\cal B}) d \,\tilde{c}({\cal B})^2,\quad \tilde{c}({\cal B})=\max_J{d_J({\cal B})/n_J({\cal B})}
$$ where we used (\ref{eq:proj-basis}).
By defining $ c({\cal B}):=2\,\sqrt{ 2\, d({\cal B})}\, \tilde{c}({\cal B})$) and bringing everything together  one obtains
\begin{eqnarray}
\mathbb{E}_{U,V}\left[d_{{\cal B},\rho_0}(U, V)\right]\le 2\sqrt{2\sum_{\beta, \gamma} \bar{b}_\beta b_\gamma C_{\beta\gamma}}\le
2
 \|b||\sqrt{2 d}\, S^{1/2}({\cal A}: {\cal B})\le    d\,  {c}({\cal B}) \,S^{1/2}({\cal A}: {\cal B}).
\end{eqnarray}
Now,  we complete the proof  by  exploiting the elementary Markov inequality
\begin{eqnarray}
\mathrm{Pr}\left[ d_{{\cal B},\rho_0}(U, V)\ge  \epsilon  \right]\le  \epsilon^{-1}  \mathbb{E}_{U,V}\left[ d_{{\cal B},\rho_0}(U, V)\right]
\le  \frac{d}{\epsilon}\,  c({\cal B})\,S^{1/2}({\cal A}:{\cal B}).
\end{eqnarray}


\section{Proof of Prop.2}
{\em{i)}}  First,  form Eq.~(\ref{eq:swap-lin}) one gets
\begin{eqnarray}
 \frac{1}{2d}\|[U, V]\|_2^2=1-\frac{1}{d}\mathrm{Re\,Tr}\left( UVU^\dagger V^\dagger\right)=
 1-\frac{1}{d}\mathrm{Re\,Tr}\left( S (U\otimes U^\dagger) (V\otimes V^\dagger)\right).
\end{eqnarray}
By Haar-averaging over $U$ $V$ one obtains Eq.~(\ref{eq:Omega}) with $\Omega_{\cal A}:=\mathbb{E}_{U\in{\cal A}}\left[U\otimes U^\dagger\right]$ and 
$ \Omega_{\cal B}:=\mathbb{E}_{U\in{\cal B}}\left[V\otimes V^\dagger\right].$
In the light of the Hilbert space decomposition (\ref{eq:hilb-decom}) one can write $U=\sum_{J}\sum_{l,m=1}^{d_J} U^J_{lm} \,\mathbf{1}_{n_J}\otimes |l\rangle\langle m|$
where the $(U^J_{lm})_{lm}$'s $(J=1,\ldots, d_Z$)  are (independent) Haar-distributed unitary matrices.  An analog expression holds for $\Omega_{\cal B}$.
Therefore
\begin{align}\label{eq:Omega_A}
\Omega_{\cal A} = \sum_{J, K}\sum_{l,m,p,q}    \mathbb{E}_{U\in{\cal A}}\left[ U^J_{lm}    \bar{U^K}_{pq} \right]   \,\mathbf{1}_{n_J}\otimes |l\rangle\langle m|\otimes
 \mathbf{1}_{n_K}\otimes |q\rangle\langle p|
 = \sum_{J}\frac{1}{d_J}\sum_{l,m}  (\mathbf{1}_{n_J}\otimes |l\rangle\langle m|)\otimes(\mathbf{1}_{n_J}\otimes |m\rangle\langle l|),
\end{align}
where we have used the basic Haar relation $$ \mathbb{E}_{U\in{\cal A}}\left[ U^J_{lm}    \bar{U^K}_{pq} \right]=\frac{1}{d_J}\delta_{J,K}\delta_{l,p}\delta_{m,q}.$$
By defining $e_\alpha:= d_J^{-1/2}   \mathbf{1}_{n_J}\otimes |l\rangle\langle m|,\,\alpha:=(J, l, m)$ one sees that $\Omega_{\cal A} =\sum_\alpha e_\alpha \otimes e_\alpha^\dagger.$
A completely analogous computation for $\Omega_{\cal B}=\sum_\beta f_\beta\otimes f_\beta^\dagger,$  completes the proof of the first line in Eq.~(\ref{eq:Omega}).
Now,  $\mathrm{Tr}\left( S\Omega_{\cal A} \Omega_{\cal B}  \right)=\sum_{\alpha, \beta} \mathrm{Tr}\left( S\, e_\alpha f_\beta\otimes e_\alpha^\dagger f_\beta^\dagger\right)=\sum_{\alpha, \beta} \mathrm{Tr}\left(e_\alpha f_\beta e_\alpha^\dagger f_\beta^\dagger\right)=
\frac{1}{2}\sum_{\alpha, \beta}
\left( \mathrm{Tr} (e_\alpha f_\beta f_\beta^\dagger e_\alpha^\dagger +  f_\beta e_\alpha e_\alpha^\dagger f_\beta^\dagger)  - \| [e_\alpha, f_\beta]\|_2^2  \right)=d -\frac{1}{2} \sum_{\alpha, \beta}  \| [e_\alpha, f_\beta]\|_2^2.$ Here we used the fact that the bases $\{e_\alpha\}_\alpha$ and $\{f_\beta\}$
are hermitean-closed and that $\sum_\alpha e_\alpha e_\alpha^\dagger=
\sum_\beta f_\beta f_\beta^\dagger=\mathbf{1}.$ From this the the second  line in Eq.~(\ref{eq:Omega}) follows.

\vspace{0.2truecm}

{\em{ii)}}
Using the definition of the algebra projection CP-maps (\ref{eq:proj-basis}) one gets that $\mathrm{Tr}\left(S\Omega_{\cal A} \Omega_{\cal B}  \right)$ is equal to
\begin{eqnarray}
\sum_{\alpha\beta} \mathrm{Tr}\left( e_\alpha f_\beta e^\dagger_\alpha f^\dagger_\beta \right)=
\sum_\alpha\mathrm{Tr}\left( e_\alpha \sum_\beta f_\beta e_\alpha^\dagger f_\beta^\dagger\right)=
\sum_\alpha \mathrm{Tr}\left( e_\alpha \mathbb{P}_{{\cal B}^\prime}(e_\alpha^\dagger)\right)=\sum_\alpha\|   \mathbb{P}_{{\cal B}^\prime}(e_\alpha)   \|_2^2.
\end{eqnarray}
Therefore $S({\cal A}:{\cal B})= 1-\frac{1}{d}\sum_\alpha\|   \mathbb{P}_{{\cal B}^\prime}(e_\alpha)   \|_2^2=\frac{1}{d}\sum_\alpha\left( \|e_\alpha\|^2 - \|   \mathbb{P}_{{\cal B}^\prime}(e_\alpha)   \|_2^2  \right)= \frac{1}{d}\sum_\alpha \|(\mathbf{1}-  \mathbb{P}_{{\cal B}^\prime})(e_\alpha)\|_2^2.$
Here we used $$\sum_\alpha\|e_\alpha\|^2=\sum_\alpha \mathrm{Tr}\left( e_\alpha e_\alpha^\dagger\right)= \mathrm{Tr}\left(\mathbb{P}_{{\cal B}^\prime}(\mathbf{1})\right)=d.$$
\vspace{0.2truecm}

{\em{iii)}} If ${\cal B}_1\subset {\cal B}_2$ then ${\cal B}^\prime_1\supset {\cal B}^\prime_2 \Rightarrow \mathbb{P}_{{\cal B}_1^\prime}\ge  \mathbb{P}_{{\cal B}_2^\prime}\Rightarrow
\mathbb{Q}_{{\cal B}_1^\prime}\le  \mathbb{Q}_{{\cal B}_2^\prime}\Rightarrow \| \mathbb{Q}_{{\cal B}_1^\prime}(e_\alpha)\|_2^2\le\| \mathbb{Q}_{{\cal B}_2^\prime}(e_\alpha) \|_2^2,\,(\forall \alpha)$ The monotonicity of $S$ then follows from Eq.~(\ref{eq:Q}).  Moreover the result for $S_2$ [Eq.~(\ref{eq:P-S_2})] follows from the fact that $-\log$ is non-increasing.
\vspace{0.2truecm}

{\em{iv)}}
From iii) one has $ S({\cal A}: {\cal B})\le S({\cal A}: L({\cal H}))=1 -\frac{1}{d}\mathrm{Tr}\left( S\Omega_{\cal A} \Omega_{L({\cal H}) }    \right) =
1-  \frac{1}{d^2}\mathrm{Tr}\,\Omega_{\cal A},$ where we used $\Omega_{L({\cal H}) }= S/d$ [See Eq.~(\ref{eq:Omega_A})].
Now,  $\mathrm{Tr}\,\Omega_{\cal A}=\sum_\alpha |\mathrm{Tr}\, e_\alpha|^2=\mathrm{Tr}_{HS}(\mathbb{P}_{{\cal A}^\prime})=d({\cal A}^\prime).$
Exchanging $\cal A$ and $\cal B$ and taking the minimum of the two upper bounds proves  (\ref{eq:upper-bound}) and from it (\ref{eq:upper-bound-2}) follows immediately,.
$\hfill\Box$

\section{Proof of Eq.~(\ref{eq:averaged-MAN})}
If $X,Y\in L({\cal H})$   and ${\cal U}^{\otimes 2}(X):=U^{\otimes 2}Y (U^\dagger)^{\otimes 2},$ by performing an Haar average over the unitary $U$ one gets 
$
\mathbb{E}_U\left[ \langle X,\, {\cal U}^{\otimes 2}(Y) \rangle    \right]=\langle X,\,\mathbb{P}(X)\rangle,
$
where the super-projection $\mathbb{P}$ onto the permutation algebra $\mathbf{C}\{\mathbf{1}, S  \}\subset L({\cal H}^{\otimes\,2})$ is given by
\begin{align}
\mathbb{P}(X):=\mathbb{E}_U\left[{\cal U}^{\otimes 2}(X)  \right]=\sum_{\alpha=\pm} \frac{\mathbf{1}+\alpha S}{\sqrt{2 d(d+\alpha)}} \langle  \frac{\mathbf{1}+\alpha S}{\sqrt{2 d(d+\alpha)}},\,  X\rangle.
\end{align}
It follows that $
\mathbb{E}_U\left[ \langle X,\, {\cal U}^{\otimes 2}(Y) \rangle    \right]=\sum_{\alpha=\pm} \langle X,  \frac{\mathbf{1}+\alpha S}{\sqrt{2 d(d+\alpha)}} \rangle\langle  \frac{\mathbf{1}+\alpha S}{\sqrt{2 d(d+\alpha)}} ,\, Y\rangle,
$
whence
\begin{eqnarray}
 \mathbb{E}_U\left[   \mathrm{Tr}\left(S\,\Omega_{\cal A}{\cal U}^{\otimes \,2}(( \Omega_{\cal B})    \right)\right]=\sum_{\alpha=\pm}
\alpha \frac{( d+\alpha d({\cal A}^\prime))(d+\alpha d({\cal B}^\prime))}{2\,d (d+\alpha)}=
\frac{d^2(d({\cal A}^\prime)+d({\cal B}^\prime)-1)-  d({\cal A}^\prime)  d({\cal B}^\prime)}{d (d^2-1)}\nonumber  
\end{eqnarray}
  where we have used that $\mathrm{Tr}(S\,\Omega_{\cal X})=d, $ and $\mathrm{Tr}(\Omega_{\cal X})=d({\cal X}^\prime)$ for ${\cal X}={\cal A},\,{\cal B}.$
  By straightforward algebra one gets
  \begin{align}\label{eq:Escher-Dec-27-2023}
 \mathbb{E}_U\left[ S({\cal A}:{\cal U}({\cal B}))\right]=1-\frac{1}{d}\mathrm{Tr}\left( S\,\Omega_{\cal A} {\cal U}^{\otimes\,2}(\Omega_{\cal B})    \right)=
\frac{\left( 1-\frac{d({\cal A}^\prime)}{d^2}\right)\left(1-\frac{d({\cal B}^\prime)}{d^2}\right)}{1-\frac{1}{d^2} }.
  \end{align}
Since $S(L({\cal H}):L({\cal H}))=1-\frac{1}{d^2},$ then,  Eq.~(\ref{eq:averaged-MAN})  follows immediately from Eqs. ~(\ref{eq:fraction})  and   (\ref{eq:Escher-Dec-27-2023}).

 Note,  this proof is fundamentally the same as the one of  Eq.~(13) in  Ref.~\cite{A-OTOC-Faidon-2022} for $\cal A$-OTOCs.
 The latter is obtained from (\ref{eq:averaged-MAN}) by setting ${\cal B}={\cal A}^\prime.$
  
\section{Proof of Eq.~(\ref{eq:Quantumness-upper-bound})}\label{sec:proof-quantumness-bound}
Let $A=A^\dagger=\sum_k \alpha_k \tilde{\Pi}_k\in{\cal A}_{\tilde{B}}$ then $\sum_i \sigma_i^2(A)=\|A\|_2^2\, \langle \mathbf{a}, (\mathbf{1}- \hat{C}) \mathbf{a}\rangle,$ where $\mathbf{a}$ is the $d$-dimensional normalized real vector with components $a_i:=\alpha_i/\|A\|_2 \,(i=1,\ldots,d)$ (note $\|A\|_2^2=\sum_k \alpha_k^2$) and the hermitean  matrix $\hat{C}$ is defined as $\hat{C}_{hk}:=\sum_{i=1}^d  \mathrm{Tr}\left( \Pi_i \tilde{\Pi}_h\right)  \mathrm{Tr}\left( \Pi_i \tilde{\Pi}_k\right)=: (X^T X)_{hk}
,\, (h,k=1,\ldots, d).$ Therefore,
\begin{eqnarray}\label{eq:Qbound}
Q_{B}(\tilde{B})=\sup_{\|A\|_2=1} \frac{1}{d}\sum_i \sigma_i^2(A)\le \frac{1}{d}\| \mathbf{1}- \hat{C}  \|_\infty\le \frac{1}{d}\| \mathbf{1}- \hat{C}  \|_1=
 \frac{1}{d}\mathrm{Tr}\left( \mathbf{1} - X^T X\right)
= 1 -\frac{1}{d} \|X|_2^2.
\end{eqnarray}
Here we used that standard operator norm inequalities and the fact that $\|Y\|_1=\mathrm{Tr}\,Y$ for an positive semi-definite matrix $Y$
(note $\langle v,  X^T X v\rangle=\|Xv\|^2\le 1$ for a normalized vector $v$ as $X$ is bistochastic,  whence $\mathbf{1}-\hat{C}\ge 0.$)
Now,  since $\|X|_2^2=\sum_{i,j}|\langle i| \tilde{j}\rangle|^4$ from (\ref{eq:MAN-MASA}) and (\ref{eq:Qbound})  one gets the first inequality in 
 Eq.~(\ref{eq:Quantumness-upper-bound})  is   complete. 

To show the second inequality,  we first note that $\sigma_i^2(A)=\frac{1}{2}\|[\Pi_i, A]\|_2^2.$ It follows that  the relative quantumness (\ref{eq:Quantumness}) can be written in terms of commutators of the projections generating the two algebras
\begin{align}
Q_B(\tilde{B})=\frac{1}{2d} \sup_{A\in {\cal A}_{\tilde{B}}, \,\|A\|_2=1} \sum_{i=1}^d \|[\Pi_i, A]\|_2^2
\end{align}
Now,  since the $\tilde{\Pi}_j$'s are in ${\cal A}_{\tilde{B}}$ and $\|\tilde{\Pi}_j\|_2=1\, (\forall j),$ one can write $$S({\cal A}_B: {\cal A}_{\tilde{B}})=\frac{1}{2d} \sum_{i,j=1}^d \|  [\Pi_i,\,\tilde{\Pi}_j   \|_2^2\le \sum_{j=1}^d\left( 
\frac{1}{2d} \sum_{i=1}^d  \|  [\Pi_i,\,\tilde{\Pi}_j   \|_2^2 \right)\le d \,Q_B(\tilde{B}).
$$

\section{Proof of Prop.  3}
By extension of $\Omega_{L({\cal H})}= S/d$ (see above) one sees that
$\Omega_{{\cal A}_S}= T_S/d^{|S|}
$  where $T_S$ swap operator in the  space ${\cal H}_\Lambda^{\otimes 2}=(\otimes {\cal H}_{i\in\Lambda})^{\otimes 2}$ associated with the region $S$ 
\begin{align}\label{eq:T_S}
T_S := (\otimes_{i\in S^c}\mathbf{1}_i)\otimes (\otimes_{i\in S} T_i),
\end{align}
 here the $T_i$'s are the swaps of the individual ${\cal H}_i$ factors in the doubled space.
 From (\ref{eq:T_S}) follows $\mathrm{Tr}(T_S)=d^{|S|} d^{2|S^c|}=d^{2|\Lambda|-|S|}.$
 Moreover,  since $T_i^2=\mathbf{1}$ the following ``swap algebra" \cite{zanardiQuantumCoherenceGenerating2018} holds $T_{S_1} T_{S_2}=T_{S_1\Delta S_2}$ where
 $S_1\Delta S_2:= (S_1\cap S_2^c)\cup(S_2\cap S_1^c)$ is the symmetric set difference.  Therefore
 \begin{align}
 \frac{1}{d^{|\Lambda|}}\mathrm{Tr}\left(  S\Omega_{S_1}  \Omega_{S_2}   \right)=\frac{ \mathrm{Tr}\left(  T_\Lambda T_{S_1} T_{S_2}  \right)}{d^{|\Lambda|+|S_1|+|S_2|}}=
 \frac{ \mathrm{Tr}\left(  T_{(S_1\Delta S_2)^c}  \right)}{d^{|\Lambda|+|S_1|+|S_2|}} =\frac{d^{| (S_1\Delta S_2)^c|}   d^{2 |S_1\Delta S_2|}}{d^{|\Lambda|+|S_1|+|S_2|}}=d^{-2\,|S_1\cap S_2|}
 \end{align}
 where we have used $|S_1\Delta S_2|=|S_1|+|S_2|-2\,|S_1\cap S_2|$ and $| (S_1\Delta S_2)^c|=|\Lambda|-|S_1\Delta S_2|. $
This proves Eq.~(\ref{eq:S-loc}),  then (\ref{eq:S_2-loc}) and (\ref{eq:NC-loc}) follow immediately. $\hfill\Box$ 

 \section{Proof of Prop.  4}
i) We start by noticing that Eq.~(\ref{eq:Omega_A})) can be written as
\begin{align}\label{eq:Omega_S-swap}
\Omega_{\cal A}=\sum_J \frac{1}{d_J} \mathbf{1}_{n_J}^{\otimes 2}\otimes S_{d_J},
\end{align}
where $S_{d_J}$ is the swap between the send factors in ${\cal H}_J^{\otimes\,2}=(\mathbf{C}^{n_J}\otimes\mathbf{C}^{d_J})^{\otimes\,2}.$
Whence, $$\mathrm{Tr}\left(  S\, \Omega_{\cal A}^2   \right)=\sum_J \frac{1}{d_J^2}\mathrm{Tr}\left(  S   \mathbf{1}_{n_J}^{\otimes 2}\otimes  \mathbf{1}_{d_J}^{\otimes 2}  \right)=\sum_J \frac{1}{d_J^2}\mathrm{Tr}\left( ( \mathbf{1}_{n_J}\otimes  \mathbf{1}_{d_J})^2    \right)=
\sum_J \frac{n_J d_J}{d_J^2}=\sum_J \frac{n_J}{d_J}.$$  
\vspace{0.2truecm}

ii) From this, $NC({\cal A})=1-\frac{1}{d}\mathrm{Tr}\left(  S\, \Omega_{\cal A}^2   \right)=\sum_Jp_J (1-1/d_J^2),$
where the $p_J:=n_J d_J/d$'s comprise a probability distribution (as $\sum_J n_J d_J=d$).  It follows that  $$NC({\cal A})\le \max_J\{1-\frac{1}{d^2_J}\} \sum_J p_J= 1-\frac{1}{\max_J d_J^2}\le 1-\frac{1}{d^2}
=NC(L({\cal H})).$$ 
\vspace{0.2truecm}

iii) If ${\cal A}$ is collinear oen has that $d({\cal A}^\prime)=\sum_Jn_J^2=\sum_J d_J^2 (n_J/d_J)^2=\lambda \sum_Jd_J^2= \lambda^2 d({\cal A})$ where
$\lambda:=\frac{n_J}{d_J}=\sqrt{\frac{d({\cal A}^\prime)}{  d({\cal A}) }},\,(\forall J),$ then
\begin{eqnarray}
NC({\cal A})=1-\frac{1}{d}\sum_J\frac{n_J}{d_J}=1-\sqrt{\frac{d({\cal A}^\prime)}{ d^2\, d({\cal A})}}\sum_J 1
=1-\frac{Z}{ d({\cal A})},
\end{eqnarray}
where we have used  $ d({\cal A}^\prime) d({\cal A})=d^2,$ (collinearity) and  $Z:=d( {\cal Z}({\cal A})$.
\section{Proof of Prop.  5}
i) Since $S=\sum_J S_{n_J}\otimes S_{d_J} $ one has that $S\,\Omega_{\cal A}$ can be written as
\begin{eqnarray}
 \sum_J \frac{1}{d_J} \mathbf{S}_{n_J}\otimes \mathbf{1}^{\otimes 2}_{d_J}=
\sqrt{\frac{d({\cal A}^\prime)}{d({\cal A})}}\sum_J \frac{1}{n_J} \mathbf{S}_{n_J}\otimes \mathbf{1}^{\otimes 2}
=\sqrt{\frac{d({\cal A}^\prime)}{d({\cal A})}} \,\Omega_{{\cal A}^\prime}
=\sqrt{\frac{d({\cal A}^\prime)}{d({\cal A})}} \sum_\alpha \tilde{e}_\alpha\otimes \tilde{e}_\alpha.\nonumber 
\end{eqnarray}
Using this expression and collinearity again one finds
\begin{eqnarray}
& &
\frac{1}{d}\mathrm{Tr}\left( S\,\Omega_{\cal A} \Omega_{{\cal B}}  \right) =\frac{1}{d({\cal A})}\, 
\mathrm{Tr}\left( \Omega_{{\cal A}^\prime} \Omega_{{\cal B}}  \right)=
\frac{1}{d({\cal A})}\, \sum_{\alpha\beta}|\mathrm{Tr}\left( \tilde{e}_\alpha f_\beta\right)|^2
=\frac{1}{d({\cal A})}\mathrm{Tr}_{HS}\left( \mathbb{P}_{\cal A} \mathbb{P}_{{\cal B}^\prime}    \right).\nonumber 
\end{eqnarray}
where we used (\ref{eq:proj-basis}) and the fact that,  if ${\cal T}(X)=\sum_i A_i X A_i^\dagger,$ then $\mathrm{Tr}_{HS}({\cal T})=\sum_i |\mathrm{Tr} A_i|^2.$
\vspace{0.2truecm}

ii) $\| \mathbb{P}_{\cal A} -\mathbb{P}_{{\cal B}^\prime} \|^2_{HS}= \| \mathbb{P}_{\cal A} \|^2_{HS}+\| \mathbb{P}_{{\cal B}^\prime}\|^2_{HS}-2\, \langle  \mathbb{P}_{\cal A},   \mathbb{P}_{{\cal B}^\prime}\rangle_{HS}.$ Since the first two terms are equal to $d({\cal A})$ the result (\ref{eq:self-MAN-dist} ) follows from the point i) above.

\vspace{0.2truecm}
iii)  Given  two projectors  $P$ and $Q$ 
one can write $P=P^\perp + P^{int},$ and $ Q=Q^\perp + P^{jnt},$ Where, $P^{int}$ projects on the intersection of the ranges of $P$ and $Q$
and $P^{\perp}$ ($Q^{\perp}$) is the projector  onto the orthocomplement of this intersection in the range of $P$ ($Q$). Whence,  $PQ= P^{int} + P^{\perp} Q^{\perp},$
and $\mathrm{Tr}\left( P Q\right)= \mathrm{Tr}\left( P^{int}\right) +\mathrm{Tr}\left( P^{\perp} Q^{\perp}\right)\ge  \mathrm{Tr}\left( P^{int}\right).$

Applying this general result one finds  $\mathrm{Tr}_{HS}\left(  \mathbb{P}_{\cal A} \mathbb{P}_{{\cal B}^\prime}    \right)\ge \mathrm{Tr}_{HS}\left( \mathbb{P}_{  {\cal A}\cap {\cal B}^\prime  }   \right) $ from which the bounds  (\ref{eq:upper-bound-intersec}) follows immediately.  
$\hfill\Box$

\section{Proof of Prop.  6} i) Given the ``singlet"
$|\Phi^+\rangle:= \frac{1}{\sqrt{d}}\sum_{i=1} |i\rangle^{\otimes\,2},$ and the CP-map $\cal T$ one defines the CJ associated state  by
$\omega({\cal T}):=({\cal T}\otimes\mathbf{1})|\Phi^+\rangle\langle\Phi^+|=\sum_{i,j=1}^d =\frac{1}{d}{\cal T}(|i\rangle\langle j|)\otimes |i\rangle\langle j|.$
Therefore, if $\cal F$ is another CP-map
\begin{eqnarray}
\langle \omega({\cal T}), \,\omega({\cal F})\rangle=\mathrm{Tr}\left(S \,\omega({\cal T})\otimes \omega({\cal F})\right)= 
 \frac{1}{d^2}\sum_{i,j=1}^d \langle {\cal T}(|i\rangle\langle j|), \, {\cal F}(|i\rangle\langle j|)\rangle=\frac{1}{d^2}
\langle {\cal T}, \,{\cal F}\rangle_{HS}. \nonumber
\end{eqnarray}
Now,  applying this to the algebra states, using Eqs.~(\ref{eq:HS-scalar-product}), Eq.~(\ref{eq:MAN-Proj}),  and noticing that
$ d({\cal A})= \mathrm{Tr}_{HS}\, \mathbb{P}_{\cal A}=\|\mathbb{P}_{\cal A}\|_{HS}^2=d^{2}\|\omega({\cal A})\|_2^2,$
 one gets the result (\ref{eq:MAN-choi}).

ii) From i) by setting ${\cal A}={\cal B}$.
$\hfill\Box$

\section{Proof of Prop.  7}
The following Lemma holds 
\begin{align}\label{eq:HS-scalar-product-op}
\langle {\cal T},\,{\cal F}\rangle_{HS}=d(d+1)\,\mathbb{E}_\phi\left[  \langle {\cal T}(\hat{\phi}),\,     {\cal F}(\hat{\phi})\rangle  \right] -d,
\end{align}
where $\hat{\phi}:=|\phi\rangle\langle\phi|$ are one-dimensional projector and the expectation is with respect to the Haar induced distribution of the $|\phi\rangle.$ 
Indeed  the swap can be expressed by an Haar average of pure states in a doubled space \cite{zanardi2001entanglement} 
\begin{align}\label{eq:swap-average}
S=d(d+1)\mathbb{E}_\phi\left[  |\phi\rangle\langle\phi|^{\otimes\,2}  \right]-\mathbf{1}^{\otimes\,2}.
\end{align}
 Plugging this result in second $S$ in Eq.~(\ref{eq:HS-scalar-product-S}),  using unitality of the maps and Eq.~(\ref{eq:swap-lin}) again proves the lemma.

From (\ref{eq:HS-scalar-product-op} )and Eq.~(\ref{eq:MAN-Proj}) the desired results  (\ref{eq:MAN-op})  and (\ref{eq:self-MAN-op}) follow immediately,
$\hfill\Box$

\section{MAN is ```Morally" Entropy}\label{sec:morally}
We begin by writing $\sum_{\beta=1}^{d({\cal B})}\|\mathbb{P}_{{\cal A}^\prime}(f_\beta)\|_2^2=\sum_{\beta=1}^{d({\cal B})}
\mathrm{Tr}\left(S\, \mathbb{P}_{{\cal A}^\prime}^{\otimes\,2}( f_\beta\otimes f_\beta^\dagger)\right)=\mathrm{Tr}\left(S\, \mathbb{P}_{{\cal A}^\prime}^{\otimes\,2}(\Omega_{\cal B})\right).$
Now we use (\ref{eq:Omega_S-swap}) to write $\Omega_{\cal B}=\sum_J \frac{1}{d_J} \mathbf{1}_{n_J}^{\otimes 2}\otimes S_{d_J}$,  and  (\ref{eq:swap-average}) 
$ S_{d_J}=d_J(d_J+1)\mathbb{E}_{\phi_J}\left[  |\phi_J\rangle\langle\phi_J|^{\otimes\,2}  \right]-\mathbf{1}_{d_J}^{\otimes\,2}.$ Therefore
\begin{align}\label{eq:MAN-is-entropy-Pur}
\mathrm{Tr}\left(S\, \mathbb{P}_{{\cal A}^\prime}^{\otimes\,2}(\Omega_{\cal B})\right)=\sum_J(d_J+1)\, n_J^2\,\mathbb{E}_{\phi_J}\left[   \| \mathbb{P}_{{\cal A}^\prime}( \frac{\mathbf{1}_{n_J}}{n_J}\otimes |\phi_J\rangle\langle\phi_J|) \|_2^2\right]-\sum_J n_J^2\,d_J \|  \mathbb{P}_{{\cal A}^\prime} (\frac{\Pi_J}{n_J d_J}) \|_2^2.
\end{align}
where $\Pi_J= \mathbf{1}_{n_J}\otimes \mathbf{1}_{d_J}$ are the central projections of $\cal B$,
Expressing the purities in terms of linear entropie i.e., $\|\rho\|_2^2=1-S_{lin}(\rho)$  and using Eq.~(\ref{eq:Q}) one gets
\begin{align}\label{eq:MAN-is-entropy}
S({\cal A}: {\cal B})=\sum_Jp_J    \Big\lbrace a_J\, \mathbb{E}_{\phi_J}\left[  S_{lin}\left(\mathbb{P}_{{\cal A}^\prime}( \frac{\mathbf{1}_{n_J}}{n_J}\otimes |\phi_J\rangle\langle\phi_J|)  \right)     \right]      - n_J\,S_{lin}\left(   \mathbb{P}_{{\cal A}^\prime}(\frac{\Pi_J}{n_J d_J})  \right) + b_J  \Big\rbrace 
\end{align}
where $p_J:=\frac{n_J d_J}{d},\,  a_J:= \frac{n_J(d_J+1)}{d_J},\, b_J:=1-\frac{n_J}{d_J}.$
By defining  the family of CP-maps ${\cal T}^J_{\cal A}: L(\mathbf{C}^{d_J})\rightarrow L({\cal H}): X\mapsto   \mathbb{P}_{{\cal A}^\prime}( \frac{\mathbf{1}}{n_J}\otimes X), $
and $\hat{\phi}_J:= |\phi_J\rangle\langle\phi_J|,$ Eq. ~(\ref{eq:MAN-is-entropy}) can rewritten as 
\begin{align}\label{eq:MAN-is-entropy-CP-maps}
S({\cal A}: {\cal B})=\sum_Jp_J    \Big\lbrace a_J\, \mathbb{E}_{\phi_J}\left[  S_{lin}\left({\cal T}^J_{\cal A}( \hat{\phi}_J \right)     \right]      - n_J\,S_{lin}\left(  {\cal T}^J_{\cal A}(\frac{\mathbf{1}_{d_J}}{d_J})  \right) + b_J  \Big\rbrace 
\end{align}
 Eq.~(\ref{eq:MAN-is-entropy-CP-maps}) is a generalization of Eq.~(13) in \cite{styliaris_information_2020} and it reduces to it when $\cal A$ is a factor and
 ${\cal B}={\cal U}({\cal A}^\prime).$ Also,  Eq.~(\ref{eq:NC}) can be readily obtained from (\ref{eq:MAN-is-entropy-Pur}) by setting ${\cal B}={\cal A},$ and Eq.~(\ref{eq:MAN-MASA})
 for $\cal A$ and $B$ maximal abelian.  [In this case $n_J=d_J=1\Rightarrow a_J=2, \,b_J=0(\forall J)$ the expectation is irrelevant and the arguments of the $S_{lin}$ are the same.]
 
\end{widetext}
\end{document}